# Design of Transport Layer Based Hybrid Covert Channel Detection Engine

Anjan K [#1], Jibi Abraham [*2], Mamatha Jadhav V [#3]

[#] *Department Of Computer Science and Engineering,*

*M.S.Ramaiah Institute of Technology,*

*Bangalore,India*

[1] annjankk_msrit@in.com [3] mamsdalvi@gmail.com

[*] *College of Engineering*

*Pune, India*

[2] jibia.comp@coep.ac.in

***Abstract :*** *Computer network is unpredictable due to information warfare and is prone to various attacks. Such attacks on network compromise the most important attribute, the privacy. Most of such attacks are devised using special communication channel called ``Covert Channel''. The word ``Covert'' stands for hidden or non-transparent. Network Covert Channel is a concealed communication path within legitimate network communication that clearly violates security policies laid down. The non-transparency in covert channel is also referred to as trapdoor. A trapdoor is unintended design within legitimate communication whose motto is to leak information. Subliminal channel, a variant of covert channel works similarly except that the trapdoor is set in a cryptographic algorithm. A composition of covert channel with subliminal channel is the ``Hybrid Covert Channel''. Hybrid covert channel is homogenous or heterogeneous mixture of two or more variants of covert channels either active at same instance or at different instances of time. Detecting such malicious channel activity plays a vital role in removing threat to the legitimate network. In this paper, we present a study of multi-trapdoor covert channels and introduce design of a new detection engine for hybrid covert channel in transport layer visualized in TCP and SSL.*

**Keywords :** *Covert Channel, Subliminal Channel, Hybrid Covert Channel, Network Security, Trapdoors*

## I. INTRODUCTION

Covert channel [1] [2] [3] is a malicious conversation within a legitimate network communication. Covert channels clearly violate the security policies laid down by the network environment allowing the information leak to the unauthorized or unknown receiver. Covert channels do not have concrete definition and are scenario oriented. Covertness in these channels exhibit behaviours like multi-trapdoor and protocol hopped where in which channelling is not constrained to pair of communication entities. A fundamental covert channel can be visualised in figure 1 depicting the covert communication model employed in the covert channel with pre-shared information encoding and decoding scheme between the covert users.

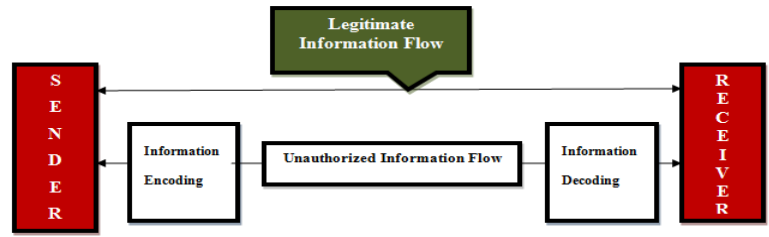

**Figure 1: Covert Channel Visualization**

Further a covert channel can exist between processes in operating system or amongst distributed objects or it can exist wherever communication is established. Focus in this paper is on exploiting the covertness in rudimentary network model. At this point of discussion, covert channel is associated with variety of similarly sounding terminologies like **side channel** or **stegnographic channel** or **supraliminal channel**. These literature terms are indifferent to each other and stand on the motto of promoting covertness in different forms or scenarios in a communication model over legitimate network.

Covert channels in general exhibit some characteristics: ***Capacity, Noise and Transmission mode*** [12]. Capacity of covert channel is the quantity of covert data that has to be transmitted. Noise is the amount of disturbance that can interfere with the covert data when transmitted in the network channel. Transmission mode is many times found to be synchronous but can also be asynchronous. A broad classification of the covert channels is described in [5].

Hybrid covert channel [5], a variant of covert channel is defined as homogeneous or heterogeneous composition of two or more covert channel variants existing either at same instance or at different instances of time. Hybrid covert channel does not have concrete composition. It is unimaginable to completely assess number of covert channels involved in hybrid composition. Hence detection is a tedious work. Complexity adds on if the hybrid covert channel behaves as multi-trapdoor and protocol hopped [9]. Hybrid covert channel here as shown in figure 2 is visualized as a combination of simple network covert channel in TCP and subliminal channel in SSL, both being transport layer protocols.

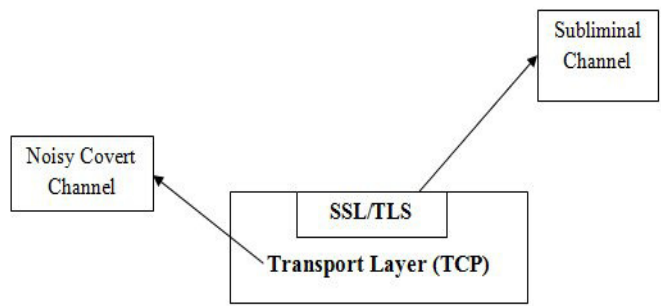

**Figure 2: Hybrid Covert Channel in Transport Layer**

Further sections of this paper cover various detection methods and design of a system to tackle the hybrid covert channel based on the proper detection method. Section II explores related work. Section III gives brief insight about various detection methodologies available in the literature. Section IV delineates about the system design. Conclusion and future enhancements are provided in section V.

## II. RELATED WORK

Extensive work has been done to devise better detection methods to detect only covert channel either on live wire or on a dataset. The method proposed in [6] is based on detecting covert shells by monitoring the unusual traffic in the network stream. Detection in covert timing channels proposed in [7] is based on packet inter-arrival and the whole process is modelled as Poisson's distribution. Illegal information flows in covert channels are tracked by tracing the Message Sequence Charts (MSC) in [8]. The paper [10] employs a statistical protocol based detection to detect hybrid covert channel based on analysis made on packet headers.

## III. DETECTION METHODS

Detection methods [10] for covert channels embedded in various protocols are relatively a new area of research. Covert channel detection is to actively monitor the illegal information flow or covert channel in the network stream. Covert Channel Identification is to identify a couple of resources used for covert channeling, especially this happen in the case of storage based covert channels. Focus in the proposed work is on active monitoring the malicious activity on the network stream and not the identification of resources. Various authors across the globe have categorized detection into following categories listed below:

### A. Signature Based Detection

It involves searching specific pre-defined patterns in the network stream and when the pattern appears, it triggers an alarm process. Best example for kind of channel it can detect is NetCat - which is a reverse-shell communication between the internal network and a public network.

### B. Protocol Based Detection

It involves searching the protocols for anomalies or violations while monitoring the network stream. This requires understanding the protocol specification described in their RFC's and detector must be knowledgeable to scan covert vulnerable fields in the protocol header. The best example for channel that can be found is Covert_TCP tool which manipulates sequence number field in TCP and IP ID in IPv4 packet for the covert communication.

### C. Behavioral Based Detection

It involves creation of user profiles and reference profiles with respect to network stream in a legitimate environment. These reference profiles are later applied to the production environment for lateral comparison of real time user profiles with reference profiles. Best instance is writing arbitrary data in any packet using stegnographic techniques.

### D. Other Approaches

Other approaches include detection based on the data mining principles like neural network and scenario based Bayes interference. Neural network approach involves training the network for `$t$' period until required accurate values to trigger the alarm process by the detection engine. In scenario based Bayes interference, a system is setup to check whether each suspicious matched signature (hypothetical attack) found in the monitored data stream is part of a global set (symptoms). Then use each global set to calculate, with a Bayes inference, the probability for a known attack to be on hold knowing the *P* (Hypothetical attack / Symptoms) probability. If the detection engine finds a suspicious scenario whose probability value is greater than a set threshold, an alarm process is triggered by the detection engine.

Above categorization can also behave as either statistical or probabilistic. A statistical approach is to run the detection engine for `*t*' hours and record an amount of data `*d*'. This period is called as learning period and such approach helps to increase the accuracy and also to set the threshold value for the alarm process. A probabilistic approach is to set a probability for the specific event *S* that occurs after the events *P*, *Q* and <u>R</u> as *y%*. This helps the detection engine to tune itself to such events in its running period.

## IV. DESIGN OF COVERT CHANNEL CREATION AND DETECTION

### A. Major Design Criteria

Hybrid covert channel is visualized here as heterogeneous combination of trapdoors placed in TCP and SSL in the transport layer. Design aspects with respect to TCP and SSL take different route. TCP packets can be captured from the network interface of a system physically connected to a small scale LAN. SSL payload is part of TCP packet. In order to detect the trapdoor in each of the protocol, first let us look at the process of formation of the TCP packet when an application data is sent from the application layer as described in figure 3.

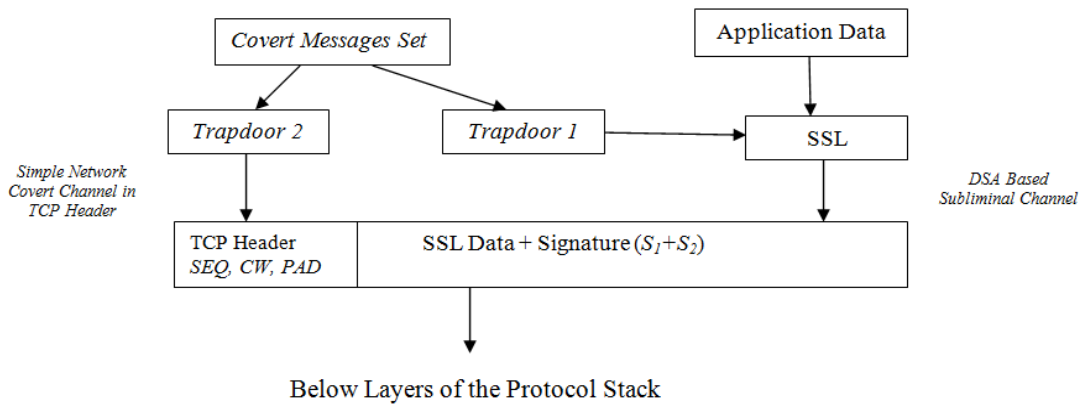

Figure 3: Hybrid Covert Channel Formation

In figure 3, words marked in italics refer to covert and those with normal font refer to legitimate process. In order to test the detection of such trapdoors in a channel, the channel itself needs to be constructed. Before going any further with design of the channels and the detection engine, a decision on the detection method to be employed plays a vital role. Network protocol like TCP is involved and hence protocol based detection can be employed but security protocol like SSL demands to have more runs to detect the trapdoor. Hence it follows a statistical approach, therefore detection method for hybrid covert channel is an amalgamation of protocol based and statistical based, termed as **Statistical Protocol Based Detection**.

**B. Designing Hybrid Covert Channel for Experimental Setup**

Design issues here takes two different routes as discussed below. With reference to figure 3, flow of design follows like first subliminal channel in SSL and then the simple network covert channel in TCP.

**1) Designing Subliminal Channel in SSL**

SSL had wide range of cipher algorithms that assist in secured communication. One such algorithm is the DSA that provide authentication service. Subliminal channel is created in DSA as per [12]. Practically this can be done in following ways

- Covert user provides his random number during the signature generation process.
- Covert user replaces system generated public-private keys with his keys.

In either of both cases, the signature component contains the subliminal activity. If private key of covert sender is known to the covert receiver then decoding is very simple. This can be programmed either with OpenSSL or JSSE secure sockets.

**2) Designing Simple Network Covert Channel in TCP**

As theoretically explained in [5] a covert sender can place his covert data in covert vulnerable fields like sequence number, Flags, Ack, options, padding and reserved. Since simple network covert channel is being constructed in this work, focus is on sequence number, padding and flags fields.

In order to implement this, a covert user needs a direct access to TCP packet generation process. Practically under a programming platform this can be implemented in two ways:
- Jpcap libraries in Java that gives direct control of the interface to the developer, here a covert user.
- BSD socket in Linux where socket creation can be done in the raw mode of operation to create custom packet and informing the kernel not to append the checksum as this is done by the developer.

**C. Design of Covert Detection Engine**

Design flow for detecting also takes two stages; one for detecting the SSL trapdoor or subliminal channel and the other for the covert channel in TCP. For TCP based covert channel, TCP packet must be available for diagnosis; this can be done by employing a protocol analyser or sniffer. For SSL, it assumed that covert user has replaced the original supplied keys and also the manipulation of random number is done. In such cases, randomness test for both keys and the random number will prove the fact that the trapdoor is placed by the covert party. Algorithm below gives a picture of the detection process.

**Algorithm for Detection Engine**

Step 1: Capture TCP packets from Network Interface from user specified network device

Step 2: Store the TCP packet.

Step 3: Analyse the TCP header on covert vulnerable fields

Step 4: Analyse the signature in TCP payload and test the key against PRNG tester

Step 5: Log the entries of the covert and subliminal activity.

Step 6: Compute the performance graph and detection content computation from the each session data set.

A single cycle of the detection engine starting with the packet capturing from the interface, then to detection and back to interface can be better understood with flow diagram depicted in figure 4.

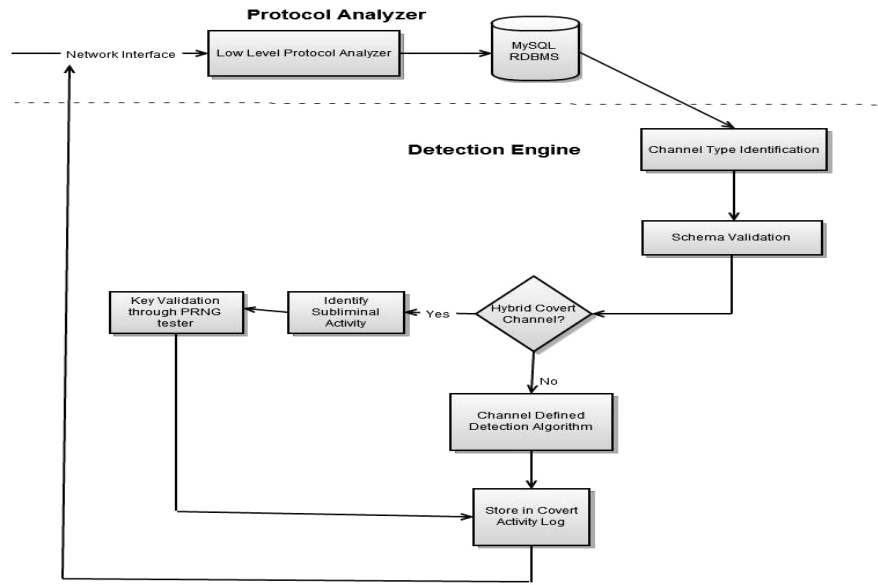

Figure 4: Detection Engine Cycle

## V. CONCLUSION AND FUTURE ENHANCEMENTS

Hard compromise on confidential information is clearly unacceptable in presence of security measures for legitimate network. Conspiracy between communication parties is not legitimate (Covert Parties) and existence of Hybrid Covert Channel is the strongest threat in communication which should be decommissioned. Conclusion is to build system to detect the activity of hybrid covert channel in a small scale LAN. This paper has focussed on designing such a system to evaluate its performance using an experimental test bed. The future enhancements include the performance evaluation of the system using the real time test bed. Also enhancements could be planned to cover most of the possible covert fields in TCP packet header like acknowledgment bounce and options.

## VI. ACKNOWLEDGEMENTS

We thank Prof. V Muralidaran, Department Head, Department of Computer Science and Engineering, M.S.Ramaiah Institute of Technology, Bangalore for his constant support and encouragement.

Anjan Koundinya thanks Late Dr. V.K Ananthashayana, Erstwhile Head, Department of Computer Science and Engineering, M.S.Ramaiah Institute of Technology, Bangalore, for igniting the passion for research.

AUTHORS PROFILE

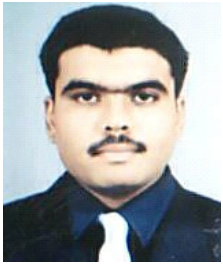

Anjan K has received his B.E degree from Visveswariah Technological University, Belgaum, India in 2007 and his M.Tech degree in Department of Computer Science and Engineering, M.S. Ramaiah Institute of Technology, Bangalore, India. He has been awarded Best Performer PG 2010 and Rank holder for his academic excellence. His areas of research includes Network Security and Cryptography, Adhoc Networks and Mobile Computing.

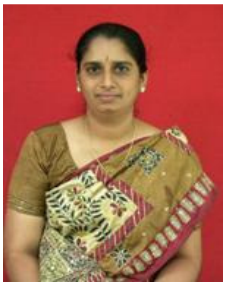

Jibi Abraham has received her M.S degree in Software Systems from BITS, Rajasthan, India in 1999 and PhD degree from Visveswariah Technological University, Belgaum, India in 2008 in the area of Network Security. Her areas of research interests include Network routing algorithms, Cryptography, Network Security of Wireless Sensor Networks and Algorithms Design. Currently she is working as a Professor at College of Engineering, Pune, India.

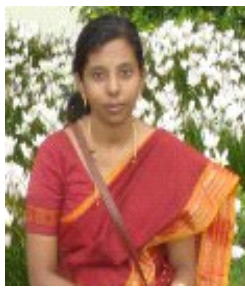

Mamatha Jadhav V has received her M.Tech degree in Computer Science and Engineering from Dr. Ambedkar Institute of Technology, Bangalore, India. She is currently working as a faculty in Department of Computer Science and Engineering, M.S. Ramaiah Institute of technology, Bangalore, India. Her subject interests include Computer Networks, DBMS and Wireless Networks.